\documentclass[aps,prl,twocolumn,amsmat,amssymb,amsfonts,superscriptaddress]{revtex4-2}
\usepackage{amsmath}
\usepackage{tikz}
\usetikzlibrary{arrows}
\usepackage{ifthen}

\usetikzlibrary{decorations.pathreplacing,decorations.markings,shapes.geometric, calc,arrows.meta}

\usepackage{times}
\usepackage{hyperref}
\usepackage{orcidlink}

\usepackage{color}
\usepackage{graphics}
\usepackage{graphicx}
\usepackage[percent]{overpic}
\usepackage{wasysym}
\usepackage{comment}

\renewcommand{\H}{\mathcal{H}}

\renewcommand{\k}{\bm{k}}

\usetikzlibrary{positioning}

\tikzset{
  on each segment/.style={
    decorate,
    decoration={
      show path construction,
      moveto code={},
      lineto code={
        \path [#1]
        (\tikzinputsegmentfirst) -- (\tikzinputsegmentlast);
      },
      curveto code={
        \path [#1] (\tikzinputsegmentfirst)
        .. controls
        (\tikzinputsegmentsupporta) and (\tikzinputsegmentsupportb)
        ..
        (\tikzinputsegmentlast);
      },
      closepath code={
        \path [#1]
        (\tikzinputsegmentfirst) -- (\tikzinputsegmentlast);
      },
    },
  },
  mid arrow/.style={postaction={decorate,decoration={
        markings,
        mark=at position .5 with {\arrow[#1]{stealth}}
      }}},
}

\def\circledarrow#1#2#3{ 
\draw[#1,->] (#2) +(80:#3) arc(80:-260:#3);
}
\def\rs{45}
\def\chrlAK#1{%
  \pgfmathparse{int(#1)}%
  \ifnum\pgfmathresult>0
  \foreach \x in {0,...,#1}
     \filldraw (1.3*\x,-.1) circle (0.1);
  \foreach \x in {0,...,#1}
     \filldraw (1.3*\x,1.1) circle (0.1);
  \draw[line width=0.2mm] (-0.2,1.1) -- (8.0,1.1);
  \draw[line width=0.2mm] (-0.2,-.1) -- (8.0,-.1);
   \pgfmathtruncatemacro{\y}{#1}
  \foreach \x in {0,...,\y}
        \draw[line width=0.2mm] (1.3*\x,1.1) -- (1.3*\x,-.1);
  \foreach \x in {0,...,5} {
     \ifthenelse{\x=0 \OR \x=2 \OR \x=4}{
     \path [draw=red,line width=\rs*0.0023882976987468mm,postaction={on each segment={mid arrow=red}}]
        (1.3*\x+0.1,0) -- (1.3*\x+1.1,0) -- (1.3*\x+0.1,1)-- (1.3*\x+0.1,0);
      \draw (1.3*\x+0.4,0.35) node {-};
      \path [draw=blue,line width=\rs*0.0023882976987468mm,postaction={on each segment={mid arrow=blue}}]
         (1.3*\x+0.2,1.0) -- (1.3*\x+1.2,1.00) -- (1.3*\x+1.2,0.00) -- (1.3*\x+0.2,1.0);
      \draw (1.3*\x+0.85,0.65) node {+};
    } {
     \path [draw=purple,line width=\rs*0.0058263516163867mm,postaction={on each segment={mid arrow=purple}}]
        (1.3*\x+0.1,0) -- (1.3*\x+1.1,0) -- (1.3*\x+0.1,1)-- (1.3*\x+0.1,0);
      \draw (1.3*\x+0.4,0.35) node {-};
      \path [draw=blue,line width=\rs*0.0058263516163867mm,postaction={on each segment={mid arrow=blue}}]
         (1.3*\x+0.2,1.0) -- (1.3*\x+1.2,1.00) -- (1.3*\x+1.2,0.00) -- (1.3*\x+0.2,1.0);
      \draw (1.3*\x+0.85,0.65) node {+};
      }
    }
    \fi
  }

\def\ra{15}
\def\chrlalpha#1{%
  \pgfmathparse{int(#1)}%
  \ifnum\pgfmathresult>0
  \foreach \x in {0,...,#1}
     \filldraw (1.3*\x,-.1) circle (0.1);
  \foreach \x in {0,...,#1}
     \filldraw (1.3*\x,1.1) circle (0.1);
  \draw[line width=0.2mm] (-0.2,1.1) -- (8.0,1.1);
  \draw[line width=0.2mm] (-0.2,-.1) -- (8.0,-.1);
   \pgfmathtruncatemacro{\y}{#1}
  \foreach \x in {0,...,\y}
        \draw[line width=0.2mm] (1.3*\x,1.1) -- (1.3*\x,-.1);
  \foreach \x in {0,...,5} {
     \ifthenelse{\x=0 \OR \x=2 \OR \x=4}{
     \path [draw=blue,line width=\ra*0.0173986550528829mm,postaction={on each segment={mid arrow=blue}}]
        (1.3*\x+0.1,0) -- (1.3*\x+0.1,1) -- (1.3*\x+1.1,0)-- (1.3*\x+0.1,0);
      \draw (1.3*\x+0.4,0.35) node {+};
      \path [draw=red,line width=\ra*0.0173986550528829mm,postaction={on each segment={mid arrow=red}}]
         (1.3*\x+0.2,1.0) -- (1.3*\x+1.2,0.00)-- (1.3*\x+1.2,1.00) -- (1.3*\x+0.2,1.0);
      \draw (1.3*\x+0.85,0.65) node {-};
    } {
     \path [draw=red,line width=\ra*0.0173986550528829mm,postaction={on each segment={mid arrow=red}}]
        (1.3*\x+0.1,0) -- (1.3*\x+1.1,0) -- (1.3*\x+0.1,1)-- (1.3*\x+0.1,0);
      \draw (1.3*\x+0.4,0.35) node {-};
      \path [draw=blue,line width=\ra*0.0173986550528829mm,postaction={on each segment={mid arrow=blue}}]
         (1.3*\x+0.2,1.0) -- (1.3*\x+1.2,1.00) -- (1.3*\x+1.2,0.00) -- (1.3*\x+0.2,1.0);
      \draw (1.3*\x+0.85,0.65) node {+};
      }
    }
    \fi
  }

\def\chrldelta#1{%
  \pgfmathparse{int(#1)}%
  \ifnum\pgfmathresult>0
  \foreach \x in {0,...,#1}
     \filldraw (1.3*\x,-.1) circle (0.1);
  \foreach \x in {0,...,#1}
     \filldraw (1.3*\x,1.1) circle (0.1);
  \draw[line width=0.2mm] (-0.2,1.1) -- (8.0,1.1);
  \draw[line width=0.2mm] (-0.2,-.1) -- (8.0,-.1);
   \pgfmathtruncatemacro{\y}{#1}
  \foreach \x in {0,...,\y}
        \draw[line width=0.2mm] (1.3*\x,1.1) -- (1.3*\x,-.1);
  \foreach \x in {0,...,5} {
     \ifthenelse{\x=0 \OR \x=2 \OR \x=4}{
     \path [draw=red,line width=\ra*0.0254952100792158mm,postaction={on each segment={mid arrow=red}}]
        (1.3*\x+0.1,0) -- (1.3*\x+1.1,0) -- (1.3*\x+0.1,1)-- (1.3*\x+0.1,0);
      \draw (1.3*\x+0.4,0.35) node {-};
      \path [draw=purple,line width=\ra*0.0372949466940812mm,postaction={on each segment={mid arrow=purple}}]
         (1.3*\x+0.2,1.0) -- (1.3*\x+1.2,0.00) -- (1.3*\x+1.2,1.00) -- (1.3*\x+0.2,1.0);
      \draw (1.3*\x+0.85,0.65) node {-};
    } {
     \path [draw=purple,line width=\ra*0.0372949466940812mm,postaction={on each segment={mid arrow=purple}}]
        (1.3*\x+0.1,0) -- (1.3*\x+1.1,0) -- (1.3*\x+0.1,1)-- (1.3*\x+0.1,0);
      \draw (1.3*\x+0.4,0.35) node {-};
      \path [draw=red,line width=\ra*0.0254952100792158mm,postaction={on each segment={mid arrow=red}}]
         (1.3*\x+0.2,1.0) -- (1.3*\x+1.2,0.00) -- (1.3*\x+1.2,1.00) -- (1.3*\x+0.2,1.0);
      \draw (1.3*\x+0.85,0.65) node {-};
      }
    }
    \fi
  }
  
\def\fournp#1{%
  \pgfmathparse{int(#1)}%
  \ifnum\pgfmathresult>0
  \foreach \x in {0,...,#1}
     \pgfmathtruncatemacro{\y}{2*\x+2}
     \draw (1.3*\x+1.3,-.1) node[circle,scale = 0.6, fill=black, black, draw, label=below:\y](b){};
  \foreach \x in {0,...,#1}
     \pgfmathtruncatemacro{\y}{2*\x+1}
     \draw (1.3*\x,1.1) node[circle,scale = 0.6, fill=black, black, draw, label=above:\y](b){};
   \pgfmathtruncatemacro{\y}{#1}
  \foreach \x in {0,...,\y} {
    \ifthenelse{\x=0 \OR \x=3 \OR \x=6}{
      \draw[line width=0.2mm] (1.3*\x,1.1) -- node[above] {$x'$} (1.3*\x+1.3,1.1);
      \ifthenelse{\x>0} {\draw[line width=0.2mm] (1.3*\x,-.1) -- node[below] {$x'$} (1.3*\x+1.3,-.1);}
      } {
    \ifthenelse{\x=1 \OR \x=4 \OR \x=7}{
      \draw[line width=0.2mm] (1.3*\x,-.1) -- node[below] {$z'$} (1.3*\x+1.3,-.1);
      \ifthenelse{\x<4} {\draw[line width=0.2mm] (1.3*\x,1.1) -- node[above] {$z'$} (1.3*\x+1.3,1.1);}
    }{
    \ifthenelse{\x=2 \OR \x=5 \OR \x=8}{
      \draw[line width=0.2mm] (1.3*\x,-.1) -- node[below] {$y'$} (1.3*\x+1.3,-.1);
      \ifthenelse{\x<\y}{\draw[line width=0.2mm] (1.3*\x,1.1) -- node[above] {$y'$} (1.3*\x+1.3,1.1);}
    }{}}
  }
  }
   \pgfmathtruncatemacro{\y}{#1}
  \foreach \x in {1,...,\y} {
    \ifthenelse{\x=0 \OR \x=3 \OR \x=6}{
      \draw[line width=0.2mm] (1.3*\x,1.1) -- node[left] {$z'$} (1.3*\x,-.1);
      \ifthenelse{\x=3}{
      \filldraw (1.3*\x-0.625,0.5) node (text) {${\color{red}\otimes}$};
      }
      } {
    \ifthenelse{\x=1 \OR \x=4 \OR \x=7}{
      \draw[line width=0.2mm] (1.3*\x,1.1) -- node[left] {$y'$} (1.3*\x,-.1);
    }{
    \ifthenelse{\x=2 \OR \x=5 \OR \x=8}{
      \draw[line width=0.2mm] (1.3*\x,1.1) -- node[left] {$x'$} (1.3*\x,-.1);
    }{}}
  }
  }
  \fi
  }

\def\slantkg4nusix#1{%
  \pgfmathparse{int(#1)}%
  \ifnum\pgfmathresult>0
  \foreach \x in {0,...,#1}
     \pgfmathtruncatemacro{\y}{2*\x+2}
     \draw (1.3*\x+1.3,-.1) node[circle,scale = 0.6, fill=black, black, draw, label=below:\y](b){};
  \foreach \x in {0,...,#1}
     \pgfmathtruncatemacro{\y}{2*\x+1}
     \draw (1.3*\x,1.1) node[circle,scale = 0.6, fill=black, black, draw, label=above:\y](b){};
   \pgfmathtruncatemacro{\y}{#1}
  \foreach \x in {0,...,\y} {
    \ifthenelse{\x=0 \OR \x=3 \OR \x=6}{
      \draw[line width=0.2mm] (1.3*\x,1.1) -- node[above] {$x'$} (1.3*\x+1.3,1.1);
      \ifthenelse{\x>0} {\draw[line width=0.2mm] (1.3*\x,-.1) -- node[below] {$x'$} (1.3*\x+1.3,-.1);}
      } {
    \ifthenelse{\x=1 \OR \x=4 \OR \x=7}{
      \draw[line width=0.2mm] (1.3*\x,1.1) -- node[above] {$z'$} (1.3*\x+1.3,1.1);
      \draw[line width=0.2mm] (1.3*\x,-.1) -- node[below] {$z'$} (1.3*\x+1.3,-.1);
    }{
    \ifthenelse{\x=2 \OR \x=5 \OR \x=8}{
      \draw[line width=0.2mm] (1.3*\x,-.1) -- node[below] {$y'$} (1.3*\x+1.3,-.1);
      \ifthenelse{\x<\y}{\draw[line width=0.2mm] (1.3*\x,1.1) -- node[above] {$y'$} (1.3*\x+1.3,1.1);}
    }{}}
  }
  }
   \pgfmathtruncatemacro{\y}{#1}
  \foreach \x in {1,...,\y} {
    \ifthenelse{\x=0 \OR \x=3 \OR \x=6}{
      \draw[line width=0.2mm] (1.3*\x,1.1) -- node[left] {$z'$} (1.3*\x,-.1);
      \ifthenelse{\x=3}{
      \filldraw (1.3*\x,0.5) node (text) {${\color{red}\otimes}$};
      }
      } {
    \ifthenelse{\x=1 \OR \x=4 \OR \x=7}{
      \draw[line width=0.2mm] (1.3*\x,1.1) -- node[left] {$y'$} (1.3*\x,-.1);
    }{
    \ifthenelse{\x=2 \OR \x=5 \OR \x=8}{
      \draw[line width=0.2mm] (1.3*\x,1.1) -- node[left] {$x'$} (1.3*\x,-.1);
    }{}}
  }
  }
  \fi
  }

\def\kgusix#1{%
  \pgfmathparse{int(#1)}%
  \ifnum\pgfmathresult>0
  \foreach \x in {0,...,#1}
     \pgfmathtruncatemacro{\y}{2*\x+1}
     \draw (1.3*\x,-.1) node[circle,scale = 0.6, fill=black, black, draw, label=below:\y](b){};
  \foreach \x in {0,...,#1}
     \pgfmathtruncatemacro{\y}{2*\x+2}
     \draw (1.3*\x,1.1) node[circle,scale = 0.6, fill=black, black, draw, label=above:\y](b){};
  \draw[line width=0.2mm,dashed] (-0.2,0.5) -- (6.7,0.5);
   \pgfmathtruncatemacro{\y}{#1-1}
  \foreach \x in {0,...,\y} {
    \ifthenelse{\x=0 \OR \x=3 \OR \x=6}{
      \draw[line width=0.2mm] (1.3*\x,1.1) -- node[above] {$x'$} (1.3*\x+1.3,1.1);
      \draw[line width=0.2mm] (1.3*\x,-.1) -- node[below] {$x'$} (1.3*\x+1.3,-.1);
      } {
    \ifthenelse{\x=1 \OR \x=4 \OR \x=7}{
      \draw[line width=0.2mm] (1.3*\x,1.1) -- node[above] {$z'$} (1.3*\x+1.3,1.1);
      \draw[line width=0.2mm] (1.3*\x,-.1) -- node[below] {$z'$} (1.3*\x+1.3,-.1);
    }{
    \ifthenelse{\x=2 \OR \x=5 \OR \x=8}{
      \draw[line width=0.2mm] (1.3*\x,1.1) -- node[above] {$y'$} (1.3*\x+1.3,1.1);
      \draw[line width=0.2mm] (1.3*\x,-.1) -- node[below] {$y'$} (1.3*\x+1.3,-.1);
      \ifthenelse{\x=2}{
      \filldraw (1.3*\x+0.7,0.5) node (text) {${\color{red}\otimes}$};
      }
    }{}}
  }
  }
   \pgfmathtruncatemacro{\y}{#1}
  \foreach \x in {0,...,\y} {
    \ifthenelse{\x=0 \OR \x=3 \OR \x=6}{
      \draw[line width=0.2mm] (1.3*\x,1.1) -- node[above left] {$z'$} (1.3*\x,-.1);
      } {
    \ifthenelse{\x=1 \OR \x=4 \OR \x=7}{
      \draw[line width=0.2mm] (1.3*\x,1.1) -- node[above left] {$y'$} (1.3*\x,-.1);
    }{
    \ifthenelse{\x=2 \OR \x=5 \OR \x=8}{
      \draw[line width=0.2mm] (1.3*\x,1.1) -- node[above left] {$x'$} (1.3*\x,-.1);
    }{}}
  }
  }
  \fi
  }

\def\wkgusixw#1{%
  \pgfmathparse{int(#1)}%
  \ifnum\pgfmathresult>0
  \foreach \x in {0,...,#1}
     \pgfmathtruncatemacro{\y}{2*\x+1}
     \draw (1.3*\x,-.1) node[circle,scale = 0.6, fill=black, black, draw, label=below:\y](b){};
  \foreach \x in {0,...,#1}
     \pgfmathtruncatemacro{\y}{2*\x+2}
     \draw (1.3*\x,1.1) node[circle,scale = 0.6, fill=black, black, draw, label=above:\y](b){};
  \draw[line width=0.2mm] (-0.2,1.1) -- (6.7,1.1);
  \draw[line width=0.2mm] (-0.2,-.1) -- (6.7,-.1);
   \pgfmathtruncatemacro{\y}{#1-1}
  \foreach \x in {0,...,\y} {
    \ifthenelse{\x=0 \OR \x=3 \OR \x=6}{
      \draw[line width=0.2mm] (1.3*\x,1.1) -- node[above] {$\Gamma_y\Gamma_z$} (1.3*\x+1.3,1.1);
      \draw[line width=0.2mm] (1.3*\x,-.1) -- node[below] {$\Gamma_y\Gamma_z$} (1.3*\x+1.3,-.1);
      \ifthenelse{\x=0}{
        \filldraw (1.3*\x+0.5,0.5) node (text) {$K^{k,l}_{i,j}$};
       \circledarrow{thick, red}{text}{0.37cm};
      }{
        \filldraw (1.3*\x+0.5,0.5) node (text) {$K^{k,l}_{i,j}$};
       \circledarrow{thick, red}{text}{0.37cm};
      }
      } {
    \ifthenelse{\x=1 \OR \x=4 \OR \x=7}{
      \draw[line width=0.2mm] (1.3*\x,1.1) -- node[above] {${\color{red} K_z}\Gamma_y$} (1.3*\x+1.3,1.1);
      \draw[line width=0.2mm] (1.3*\x,-.1) -- node[below] {${\color{red} K_z}\Gamma_y$} (1.3*\x+1.3,-.1);
      \filldraw (1.3*\x+0.5,0.5) node (text) {$\Gamma^{k,l}_{i,j}$};
      \circledarrow{thick, blue}{text}{0.37cm};
    }{
    \ifthenelse{\x=2 \OR \x=5 \OR \x=8}{
      \draw[line width=0.2mm] (1.3*\x,1.1) -- node[above] {${\color{red} K_y}\Gamma_z$} (1.3*\x+1.3,1.1);
      \draw[line width=0.2mm] (1.3*\x,-.1) -- node[below] {${\color{red} K_y}\Gamma_z$} (1.3*\x+1.3,-.1);
      \filldraw (1.3*\x+0.5,0.5) node (text) {$\Gamma^{k,l}_{i,j}$};
      \circledarrow{thick, blue}{text}{0.37cm};
    }{}}
  }
  }
   \pgfmathtruncatemacro{\y}{#1}
  \foreach \x in {0,...,\y} {
    \ifthenelse{\x=0 \OR \x=3 \OR \x=6}{
      \draw[line width=0.2mm] (1.3*\x,1.1) -- node[left] {$z'$} (1.3*\x,-.1);
      } {
    \ifthenelse{\x=1 \OR \x=4 \OR \x=7}{
      \draw[line width=0.2mm] (1.3*\x,1.1) -- node[left] {$y'$} (1.3*\x,-.1);
    }{
    \ifthenelse{\x=2 \OR \x=5 \OR \x=8}{
      \draw[line width=0.2mm] (1.3*\x,1.1) -- node[left] {$x'$} (1.3*\x,-.1);
    }{}}
  }
  }
  \fi
  }

\def\rungtripletA#1{%
  \pgfmathparse{int(#1)}%
  \ifnum\pgfmathresult>0
  \draw (-0.6,0.5) node (b){A};
  \foreach \x in {0,...,#1}
     \pgfmathtruncatemacro{\y}{2*\x+1}
     \draw (1.3*\x,-.1) node[circle,scale = 0.6, fill=gray, gray, draw](b){};
  \foreach \x in {0,...,#1}
     \pgfmathtruncatemacro{\y}{2*\x+2}
     \draw (1.3*\x,1.1) node[circle,scale = 0.6, fill=gray, gray, draw](b){};
  \draw[line width=0.2mm,gray] (-0.2,1.1) -- (8.0,1.1);
  \draw[line width=0.2mm,gray] (-0.2,-.1) -- (8.0,-.1);
   \pgfmathtruncatemacro{\y}{#1-1}
  \foreach \x in {0,...,\y} {
      \draw[line width=0.2mm,gray] (1.3*\x,1.1) -- (1.3*\x+1.3,1.1);
      \draw[line width=0.2mm,gray] (1.3*\x,-.1) -- (1.3*\x+1.3,-.1);
  }
  \pgfmathtruncatemacro{\y}{#1}
  \foreach \x in {0,...,\y} {
    \ifthenelse{\x=0 \OR \x=3 \OR \x=6}{
      \draw[line width=0.2mm,gray] (1.3*\x,1.1) --  (1.3*\x,-.1);
      } {
    \ifthenelse{\x=1 \OR \x=4 \OR \x=7}{
      \draw[line width=0.2mm,gray] (1.3*\x,1.1) -- (1.3*\x,-.1);
    }{
    \ifthenelse{\x=2 \OR \x=5 \OR \x=8}{
      \draw[line width=0.2mm,gray] (1.3*\x,1.1) --  (1.3*\x,-.1);
    }{}}
  }
  }
  \pgfmathtruncatemacro{\y}{#1}
  \foreach \x in {0,...,\y} {
    \ifthenelse{\x=0 \OR \x=1 \OR \x=6}{
        \filldraw (1.3*\x,0.5) node (text) {$t_x$};
        \filldraw[green, fill opacity=0.3] (1.3*\x,0.5) ellipse (0.24 and 0.9);
      }{}
    \ifthenelse{\x=2 \OR \x=3}{
        \filldraw (1.3*\x,0.5) node (text) {$t_y$};
        \filldraw[red, fill opacity=0.3] (1.3*\x,0.5) ellipse (0.24 and 0.9);
      }{}
    \ifthenelse{\x=4 \OR \x=5}{
        \filldraw (1.3*\x,0.5) node (text) {$t_z$};
        \filldraw[blue, fill opacity=0.3] (1.3*\x,0.5) ellipse (0.24 and 0.9);
      }{}
  }
  \fi
  }

\def\rungtripletB#1{%
  \pgfmathparse{int(#1)}%
  \ifnum\pgfmathresult>0
  \draw (-0.6,0.5) node (b){B};
  \foreach \x in {0,...,#1}
     \pgfmathtruncatemacro{\y}{2*\x+1}
     \draw (1.3*\x,-.1) node[circle,scale = 0.6, fill=gray, gray, draw](b){};
  \foreach \x in {0,...,#1}
     \pgfmathtruncatemacro{\y}{2*\x+2}
     \draw (1.3*\x,1.1) node[circle,scale = 0.6, fill=gray, gray, draw](b){};
  \draw[line width=0.2mm,gray] (-0.2,1.1) -- (8.0,1.1);
  \draw[line width=0.2mm,gray] (-0.2,-.1) -- (8.0,-.1);
   \pgfmathtruncatemacro{\y}{#1-1}
  \foreach \x in {0,...,\y} {
      \draw[line width=0.2mm,gray] (1.3*\x,1.1) -- (1.3*\x+1.3,1.1);
      \draw[line width=0.2mm,gray] (1.3*\x,-.1) -- (1.3*\x+1.3,-.1);
  }
  \pgfmathtruncatemacro{\y}{#1}
  \foreach \x in {0,...,\y} {
    \ifthenelse{\x=0 \OR \x=3 \OR \x=6}{
      \draw[line width=0.2mm,gray] (1.3*\x,1.1) --  (1.3*\x,-.1);
      } {
    \ifthenelse{\x=1 \OR \x=4 \OR \x=7}{
      \draw[line width=0.2mm,gray] (1.3*\x,1.1) -- (1.3*\x,-.1);
    }{
    \ifthenelse{\x=2 \OR \x=5 \OR \x=8}{
      \draw[line width=0.2mm,gray] (1.3*\x,1.1) --  (1.3*\x,-.1);
    }{}}
  }
  }
  \pgfmathtruncatemacro{\y}{#1}
  \foreach \x in {0,...,\y} {
    \ifthenelse{\x=0 \OR \x=5 \OR \x=6}{
        \filldraw (1.3*\x,0.5) node (text) {$t_y$};
        \filldraw[red, fill opacity=0.3] (1.3*\x,0.5) ellipse (0.24 and 0.9);
      }{}
    \ifthenelse{\x=3 \OR \x=4}{
        \filldraw (1.3*\x,0.5) node (text) {$t_x$};
        \filldraw[green, fill opacity=0.3] (1.3*\x,0.5) ellipse (0.24 and 0.9);
      }{}
    \ifthenelse{\x=1 \OR \x=2}{
        \filldraw (1.3*\x,0.5) node (text) {$t_z$};
        \filldraw[blue, fill opacity=0.3] (1.3*\x,0.5) ellipse (0.24 and 0.9);
      }{}
  }
  \fi
  }

\tikzset{
    my hex/.style={regular polygon, regular polygon sides=6, draw, inner sep=0pt, outer sep=0pt, minimum size=1cm},
    my circ/.style={draw, circle, fill=red!50!white, inner sep=0pt, minimum size=1.5mm}
}
\makeatletter
\newif\ifrainbowpolygon@rounded@outer
\newif\ifrainbowpolygon@rounded@inner
\newcounter{rainbowpolygon@colours}
\tikzset{%
  pics/rainbow polygon/.style={%
    code={%
      \tikzset{%
        rainbow polygon/.cd,
        defaults,
        #1
      }
      \ifrainbowpolygon@rounded@outer\tikzset{rainbow polygon/clip/.style={rounded corners=.5*\pgflinewidth}}\fi
      \begin{scope}[line width=\rainbowpolygon@linewidth]
        \foreach \i in {0,...,\rainbowpolygon@last}  \path ({00-\rainbowpolygon@eangle+\i*\rainbowpolygon@eangle}:\rainbowpolygon@size) coordinate (p\i) \ifrainbowpolygon@rounded@inner ++({90-\rainbowpolygon@eangle+\i*\rainbowpolygon@eangle}:-.5*\pgflinewidth) coordinate (q\i) \fi ;
         \clip [rainbow polygon/clip] (p0) \foreach \i in {1,...,\rainbowpolygon@last} { -- (p\i) } -- cycle \ifrainbowpolygon@rounded@inner (q0) \foreach \j in {\rainbowpolygon@last,...,1} { -- (q\j) } -- cycle \fi ;
         \ifrainbowpolygon@rounded@inner
           \foreach \i [evaluate=\i as \k using {int(mod(\i+1,\rainbowpolygon@sides))}, evaluate=\i as \j using {int(mod(\i,\therainbowpolygon@colours))}] in {0,...,\rainbowpolygon@last} \path [fill, col\j] (p\i) -- (p\k) -- (0,0) -- cycle;
         \else
           \pgfmathsetmacro\xyz{.5*\pgflinewidth/tan(.5*\rainbowpolygon@iangle)}
           \foreach \i [evaluate=\i as \k using {int(mod(\i+1,\rainbowpolygon@sides))}, evaluate=\i as \j using {int(mod(\i,\therainbowpolygon@colours))}] in {0,...,\rainbowpolygon@last} \draw [col\j, {Triangle Cap[length=\xyz pt]}-{Triangle Cap[length=\xyz pt]}] (p\i) -- (p\k);
         \fi
    \end{scope}
    }
  },
  rainbow polygon/.search also={/tikz},
  rainbow polygon/.cd,
  angles/.code={%
    \pgfmathsetmacro\rainbowpolygon@eangle{360/#1}%
    \pgfmathsetmacro\rainbowpolygon@iangle{180-\rainbowpolygon@eangle}%
  },
  last/.code={%
    \pgfmathsetmacro\rainbowpolygon@last{int(#1-1)}%
  },
  line width/.store in=\rainbowpolygon@linewidth,
  sides/.store in=\rainbowpolygon@sides,
  sides/.forward to=/tikz/rainbow polygon/angles,
  sides/.forward to=/tikz/rainbow polygon/last,
  size/.store in=\rainbowpolygon@size,
  rounded outer corners/.is if=rainbowpolygon@rounded@outer,
  rounded inner corners/.is if=rainbowpolygon@rounded@inner,
  rounded corners/.is choice,
  rounded corners/none/.style={/tikz/rainbow polygon/.cd, rounded outer corners=false, rounded inner corners=false},
  rounded corners/outer/.style={/tikz/rainbow polygon/rounded outer corners},
  rounded corners/inner/.style={/tikz/rainbow polygon/rounded inner corners},
  rounded corners/both/.style={/tikz/rainbow polygon/.cd, rounded outer corners, rounded inner corners},
  clip/.style={},
  colours/.code={%
    \setcounter{rainbowpolygon@colours}{0}%
    \edef\tempa{#1}%
    \foreach \i in \tempa {
      \edef\tempb{col\therainbowpolygon@colours}%
      \xglobal\colorlet{\tempb}{\i}%
      \stepcounter{rainbowpolygon@colours}%
    }%
  },
  colours/.default={magenta,blue,green,orange,red,gray},
  defaults/.style={%
    line width=5pt,
    sides=5,
    size=10mm,
    rounded corners=none,
    colours,
  },
}

\newcommand{\TITRE}{Exact Ground States and Phase Diagram of the Quantum Compass Model under an in-plane Field}
\newcommand{\SC}{$SVC$}
\newcommand{\SCX}{$Z$}
\newcommand{\SCZ}{$X$}
\newcommand{\LF}{$LR$}
\newcommand{\LFZ}{$R$}
\newcommand{\LFX}{$L$}
\newcommand{\shalf}{$S\mathord{=}1/2$}
\newcommand{\hcup}{$h_{xz}^{c2}$}
\newcommand{\hclow}{$h_{xz}^{c1}$}
\newcommand{\hcone}{0.540}
\newcommand{\hctwo}{1.626}
\definecolor{citeblue}{HTML}{3553FF}
\definecolor{cblue}{RGB}{14,107,237}
\definecolor{cred}{RGB}{228,26,28}
\definecolor{fgreen}{HTML}{228B22}
\definecolor{orange}{HTML}{CC8400}

\begin{document}

\title{\TITRE}
\author{A. D. S. Richards}
\email{richaa12@mcmaster.ca}
\affiliation{Department of Physics and Astronomy, McMaster University, Hamilton, Ontario L8S 4M1, Canada}
\author{Erik S. S{\o}rensen\,\orcidlink{0000-0002-5956-1190}}
\email{sorensen@mcmaster.ca}
\affiliation{Department of Physics and Astronomy, McMaster University, Hamilton, Ontario L8S 4M1, Canada}

\date{\today}

\begin{abstract}
We consider the square lattice \shalf\ quantum compass model (QCM) parameterized by $J_x, J_z$, under a field, $\mathbf{h}$, in the $x$-$z$ plane. 
At the special field value, $(h_x^\star,h_z^\star)$=$2S(J_x,J_z)$, we show that the QCM Hamiltonian may be written in a form such that two simple product states can be identified as exact ground-states, below a gap.
Exact excited states can also be found. The exact product states are characterized by a staggered vector chirality, attaining a non-zero value in the surrounding phase.
The resulting gapped phase, which we denote by \SC\ occupies most of the in-plane field phase diagram.
For some values of $h_x\mathord{>}h_z$ and $h_z\mathord{>}h_x$ at the edges of the phase diagram, we have found transitions between the \SC\ phase and phases of weakly-coupled Ising-chain states, \SCX\ and \SCZ.
In zero field, the QCM is known to have an emergent sub-extensive ground-state degeneracy.
As the field is increased from zero, we find that this degeneracy is partially lifted, resulting in bond-oriented spin-stripe states, \LFX\ and \LFZ, which are each separated from one another and the \SC\ phase by first-order transitions.
Our findings are important for understanding the field dependent phase diagram of materials with predominantly directionally-dependent Ising interactions.

\end{abstract}
\maketitle

Quantum compass models were first introduced as a model of orbital-orbital interactions arising from a Jahn-Teller distortion~\cite{Kugel72,Kugel73,Kugel_Khomskii_1982,Khomskii_2022}, and both classical and quantum versions
have been extensively studied~\cite{Ferrero_2003,Ulrich_2003,Mishra_2004,Nussinov_Fradkin_2005,Dorier_Becca_2005,Tanaka_2007,
Wenzel_Janke_2008,Orus_Vidal_2009_compass,You_2010,Trousselet_2010,Brzezicki_2013,Wang_2015,Nussinov_Brink_2015,Natori_2023,Khatua_2023}
with the main focus on ground-state properties of two-dimensional models. Interest in compass models intensified with the realization
that Kitaev's honeycomb model~\cite{kitaev2006anyons} with bond-directional interactions, a special case of a compass model, potentially
can be realized in materials through a super exchange mechanism~\cite{Jackeli_Khaliullin_2009}. 
In particular, iridium- and ruthenium-based systems in which ligands form edge-sharing octahedra surrounding the transition metal atoms have been proposed as materials which may realize a pseudospin Kitaev model~\cite{Jackeli_Khaliullin_2009}, with  $\alpha$-RuCl$_3$~\cite{plumb2014prb,banerjee2016proximate,banerjee2018npj}, a layered two-dimensional honeycomb material, as one of the most promising materials. This has given rise to the class of Kitaev materials~\cite{rau2016review,winter2017review,hermanns2018review,Takagi2019review,trebst2022review,Perkinsreview} that one may view as particular realizations of the broader class of quantum compass models.
For Kitaev materials, field-induced spin liquid phases are of special interest 
due to the potential presence of anyonic excitations, and intriguing results been observed 
in theoretical studies  
~\cite{zhu2018prb,nasu2018prb,liang2018prb,gohlke2018prb,lu2018spinon,
hickey2019visons,patel2019,zou2020neutral,Gordon2019,kaib2019prb,Lee2020Magnetic,Li2021,Zhang_2022,Hwang_2022,Feng_2023}
and  in recent experiments on $\alpha$-RuCl$_3$ when an out-of-plane field~\cite{Zhou2023} is applied in the $[111]$ direction,
as well as for an in-plane field~\cite{Kasahara2018,Yokoi2021,Czajka2021,Bruin2022,Czajka2022}.
The latter case is of special interest here since we show that for the closely related square lattice quantum compass model (QCM)
a twice degenerate exact ground-state below a gap can be found under an in-plane field,  inducing an extended phase
with other non-trivial phases in proximity. Here, we determine the complete in-plane field phase diagram.

The bulk of our results are focused on the QCM, and we first note a number of interesting properties of this model.
The QCM, in absence of a magnetic field, has a sub-extensive ground-state degeneracy of $2\mathord{\times}2^L$~\cite{Dorier_Becca_2005} 
and topological soliton excitations which are deconfined in one dimension~\cite{Nussinov_Ortiz_2009}. Through a duality transformation~
\cite{Nussinov_Fradkin_2005}, it has been shown that the QCM is equivalent to the Xu-Moore model, originally proposed to model interactions between $p$+$ip$ superconductor arrays~\cite{Xu_Moore_2004}. Furthermore, a duality mapping has also been established between the Xu-Moore model and the transverse-field toric code model~\cite{kitaev2003toric,Vidal_Dusuel_2009}. Consequently, a duality mapping exists between the zero field QCM and the transverse field toric code model,
and the latter model has been studied under an in-plane field~\cite{Tupitsyn_Stamp_2010} as well as a transverse field~\cite{Vidal_Dusuel_2009}.
Both classical and quantum QCM models have been studied at finite temperature~\cite{Mishra_2004,Wenzel_Janke_2008}, in both cases finding a
transition in the 2D Ising universality class to a low temperature ordered phase.
One may also note that, it has been shown in Ref.~\cite{Kamiya_Chern_2018} that two decoupled copies of the QCM can be mapped to the model of interacting Majorana fermions of Ref.~\cite{Kamiya_Chern_2018}, relevant to 3D topological insulators with proximity-induced superconductivity.
Dualities between each of these models demonstrate how properties of the QCM may be understood in several different contexts.

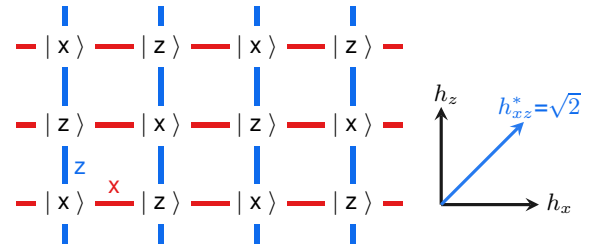
\begin{figure}[b!]
    \centering
    \vspace{-0.45cm}
    \hspace{0.05cm}
    \begin{tikzpicture}
        \node (txt1){\bf \sf $|\ $x$\ \rangle$};
        \node [right=of txt1,xshift = -0.5cm] (txt2){\bf \sf $|\ $z$\ \rangle$};
        \node [below=of txt1,yshift = 0.5cm] (txt3){\bf \sf $|\ $z$\ \rangle$};
        \node [right=of txt3, xshift = -0.5cm] (txt4){\bf \sf $|\ $x$\ \rangle$};
        \node [below=of txt3,yshift = 0.5cm] (txt5){\bf \sf $|\ $x$\ \rangle$};
        \node [below=of txt3,yshift = 0.65cm, xshift=0.65cm] {\bf \sf \color{cred} x};
        \node [below=of txt3,yshift = 0.9cm, xshift=0.2cm] {\bf \sf \color{cblue} z};
        \node [right=of txt5, xshift = -0.5cm] (txt6){\bf \sf $|\ $z$\ \rangle$};
        \node [right=of txt2,xshift = -0.5cm] (txt7){\bf \sf $|\ $x$\ \rangle$};
        \node [right=of txt7,xshift = -0.5cm] (txt8){\bf \sf $|\ $z$\ \rangle$};
        \node [below=of txt7,yshift = 0.5cm] (txt9){\bf \sf $|\ $z$\ \rangle$};
        \node [right=of txt9, xshift = -0.5cm] (txt10){\bf \sf $|\ $x$\ \rangle$};
        \node [below=of txt9,yshift = 0.5cm] (txt11){\bf \sf $|\ $x$\ \rangle$};
        \node [right=of txt11, xshift = -0.5cm] (txt12){\bf \sf $|\ $z$\ \rangle$};
        \draw[stealth-, black!90, very thick] (5.0,-0.8) --   (5.0,-2.1);
        \draw[stealth-, black!90, very thick] (6.3,-2.1) --   (5.0,-2.1);
        \draw[stealth-, cblue!90, very thick] (6.1,-1.0) --   (5.0,-2.1);
        \node [right=of txt11, xshift = 2.35cm, yshift=-0.0cm] (lbl1){\bf $h_x$};
        \node [right=of txt11, xshift = 0.85cm, yshift=1.45cm] (lbl1){\bf $h_z$};
        \node [right=of txt11, xshift = 1.7cm, yshift=1.3cm] (lbl1){\color{cblue} \bf $h^*_{xz}$=$\sqrt{2}$};
        \node [right=of txt2,xshift = -1.1cm]
        (hbnd1){\textcolor{cred}{\rule{0.5cm}{0.05cm}}};
        \node [right=of txt8,xshift = -1.1cm]
        (hbnd1){\textcolor{cred}{\rule{0.25cm}{0.055cm}}};
        \node [right=of txt7,xshift = -1.1cm]
        (hbnd1){\textcolor{cred}{\rule{0.5cm}{0.055cm}}};
        \node [right=of txt3,xshift = -1.1cm]
        (hbnd1){\textcolor{cred}{\rule{0.5cm}{0.055cm}}};
        \node [right=of txt1,xshift = -1.1cm]
        (hbnd2){\textcolor{cred}{\rule{0.5cm}{0.055cm}}};
        \node [right=of txt4,xshift = -1.1cm]
        (hbnd1){\textcolor{cred}{\rule{0.5cm}{0.055cm}}};
        \node [right=of txt5,xshift = -1.1cm]
        (hbnd1){\textcolor{cred}{\rule{0.5cm}{0.055cm}}};
        \node [right=of txt6,xshift = -1.1cm]
        (hbnd1){\textcolor{cred}{\rule{0.5cm}{0.055cm}}};
        \node [right=of txt9,xshift = -1.1cm]
        (hbnd1){\textcolor{cred}{\rule{0.5cm}{0.055cm}}};
        \node [right=of txt10,xshift = -1.1cm]
        (hbnd1){\textcolor{cred}{\rule{0.25cm}{0.055cm}}};
        \node [right=of txt11,xshift = -1.1cm]
        (hbnd2){\textcolor{cred}{\rule{0.5cm}{0.055cm}}};
        \node [right=of txt12,xshift = -1.1cm]
        (hbnd2){\textcolor{cred}{\rule{0.25cm}{0.055cm}}};
        \node [left=of txt1,xshift = 1.1cm]
        (hbnd2){\textcolor{cred}{\rule{0.25cm}{0.055cm}}};
        \node [left=of txt5,xshift = 1.1cm]
        (hbnd2){\textcolor{cred}{\rule{0.25cm}{0.055cm}}};
        \node [left=of txt3,xshift = 1.1cm]
        (hbnd2){\textcolor{cred}{\rule{0.25cm}{0.055cm}}};
        \node [below=of txt1,yshift = 1.1cm]
        (hbnd1){\textcolor{cblue}{\rule{0.055cm}{0.5cm}}};
        \node [below=of txt2,yshift = 1.1cm]
        (hbnd2){\textcolor{cblue}{\rule{0.055cm}{0.5cm}}};
        \node [below=of txt3,yshift = 1.1cm]
        (hbnd1){\textcolor{cblue}{\rule{0.055cm}{0.5cm}}};
        \node [below=of txt4,yshift = 1.1cm]
        (hbnd1){\textcolor{cblue}{\rule{0.055cm}{0.5cm}}};
        \node [below=of txt5,yshift = 1.1cm]
        (hbnd2){\textcolor{cblue}{\rule{0.055cm}{0.25cm}}};
        \node [below=of txt6,yshift = 1.1cm]
        (hbnd2){\textcolor{cblue}{\rule{0.055cm}{0.25cm}}};
        \node [below=of txt7,yshift = 1.1cm]
        (hbnd1){\textcolor{cblue}{\rule{0.055cm}{0.5cm}}};
        \node [below=of txt8,yshift = 1.1cm]
        (hbnd2){\textcolor{cblue}{\rule{0.055cm}{0.5cm}}};
        \node [below=of txt9,yshift = 1.1cm]
        (hbnd1){\textcolor{cblue}{\rule{0.055cm}{0.5cm}}};
        \node [below=of txt10,yshift = 1.1cm]
        (hbnd1){\textcolor{cblue}{\rule{0.055cm}{0.5cm}}};
        \node [below=of txt11,yshift = 1.1cm]
        (hbnd1){\textcolor{cblue}{\rule{0.055cm}{0.25cm}}};
        \node [below=of txt12,yshift = 1.1cm]
        (hbnd1){\textcolor{cblue}{\rule{0.055cm}{0.25cm}}};
        \node [above=of txt1,yshift = -1.1cm]
        (hbnd1){\textcolor{cblue}{\rule{0.055cm}{0.25cm}}};
        \node [above=of txt2,yshift = -1.1cm]
        (hbnd1){\textcolor{cblue}{\rule{0.055cm}{0.25cm}}};
        \node [above=of txt7,yshift = -1.1cm]
        (hbnd1){\textcolor{cblue}{\rule{0.055cm}{0.25cm}}};
        \node [above=of txt8,yshift = -1.1cm]
        (hbnd1){\textcolor{cblue}{\rule{0.055cm}{0.25cm}}};
    \end{tikzpicture}
    \caption{Exact ground state of the QCM under in-plane field $h^*_{xz}=2JS\sqrt{2}$. Coloured bonds represent Ising interactions.}
    \label{fig:exact}
\end{figure}
The antiferromagnetic quantum compass model is,
\begin{equation}
\H = J\sum_{\bf r}(\hat S^x_{\bf r}\hat S^x_{{\bf r}+e_x}+\hat S^z_{\bf r}\hat S^z_{{\bf r}+e_z})-\sum_{\bf r}\mathbf{h}\cdot \mathbf{\hat S_r}.
\label{eq:compass}
\end{equation}
Here, we set $g$=$\hbar$=$\mu_B$=$1$. 
Furthermore, we  parameterize the field term as
$\bm{h}$=$h(\cos\phi_{xz}\cos\theta_y,\sin\phi_{xz}\cos\theta_y,\sin\theta_y)$ and define $|\mathbf{h}|=h$ as the field strength. 
We use $N=L_x\times L_z$ to denote the number of sites in the model, and we shall refer to the $J\hat S^x\hat S^x$ coupling as a $x$-bond 
and the $J\hat S^z\hat S^z$ coupling as a $z$-bond.
In zero field, a unitary transformation around the $y$-axis on every second site, relates $J$ to $-J$. However, since our focus is on ground-states in the presence of a field, the sign of $J$ matters, and we exclusively focus on the antiferromagnetic (AF) model with $J\mathord{>}0$. We set $J\mathord{=}1$.

{\it Exact Ground and Excited States:} The exact ground-states for the QCM can be found by the following simple argument.
If we consider the Hamiltonian, Eq.~(\ref{eq:compass}) for general $S$, we can write the field term in the form $-\sum_{\bf r}(h_x\hat S^x_{\bf r}+h_z\hat S^z_{\bf r})$. Following Ref.~\cite{Sorensen2023a,Sorensen2023b}, we then see that with $\phi_{xz}$=$\pi/4$, where $h_x$=$h_z$, we can absorb the field
term into the interaction term at the special field value $h_x^\star$=$h_z^\star$=$2JS$ with $|h^\star_{xz}|$=$2JS\sqrt{2}$. For a $L_x\mathord{\times}L_z$ lattice with periodic boundary conditions in both directions and both $L_x$ and $L_z$ {\it even}, we can then write at $h^\star_{xz}$:
\begin{flalign}\label{eq:Hstar}
\ \ &\H = \H_p -2NJS^2\nonumber &\\
\ \ &\H_p = J\sum_{\bf r}\big[\left[(S-\hat S^x_{\bf r}\right)(S-\hat S^x_{{\bf r}+e_x})+\nonumber &\\
&\hskip 3cm  (S-\hat S^z_{\bf r})(S-\hat S^z_{{\bf r}+e_z})\big].
\end{flalign}
$\H_p$ is here positive semidefinite, and it follows that if a product state $|P\rangle$ can be found where each site is in an
eigenstate of $\hat S^\alpha|\alpha\rangle$=$S|\alpha\rangle$ ($\alpha$=$x,z$) such that $\H_p|P\rangle$=0, then $|P\rangle$ is not only an eigenstate, but a ground-state.
For the QCM it is straight forward to see that if $L_x$ and $L_z$ are both {\it even}, and periodic boundary conditions (PBC) are applied,
then  the two simple product states with $|x\rangle$ on one sublattice and $|z\rangle$ on the other, as shown in Fig.~\ref{fig:exact},
are eigenstates of $\H_p$ with eigenvalue 0, and therefore degenerate ground-states with $E_0$=$-2NJS^2$.
This construction trivially generalizes to the case where $J_x\mathord{\neq}J_z$ where the same ground-states appear at 
$(h_x^\star, h_z^\star)$=$2S(J_x, J_z)$. It is exact for any finite $L_x\mathord{\times} L_z$ torus under PBC, but does not hold for open boundary conditions (OBC) nor when
$L_x$ or $L_z$ are odd.
It is interesting to note that the above argument is only superficially related to the remarkable extension of the Lieb-Schultz-Mattis (LSM) theorem for quantum spin chains~\cite{Lieb1961,Affleck1986} to the case of an applied field~\cite{Oshikawa1997,Oshikawa2000},
showing that magnetization plateaus can appear, associated with a gapped state, when conserved quantities such as the total  magnetization $\sum_jS^z_j$ are present.
In contrast, for the QCM, the magnetization is not conserved, and since we can generalize to the case $J_x\mathord{\neq}J_z$, any special symmetry axis does not appear important.
We also note that similar product states formed with $S^\alpha|\alpha_m\rangle$=$m|\alpha_m\rangle$ with $0\mathord{<}m\mathord{<}S$ will be eigenstates
at the field value $h_x$=$h_z$=$2Jm$, but not ground-states. In the following, we provide strong numerical evidence for a sizable gap at $h^\star_{xz}$ and demonstrate that the
two product states are the {\it only} ground-states at $h^\star_{xz}$ under periodic boundary conditions (PBC) with $L_x$, $L_z$ even.
We expect that for large systems, lifting these constraints  will not change the physics due to the presence of a gap, and we explore
the full phase diagram using iPEPS, without imposing PBC.
\begin{figure}[t]
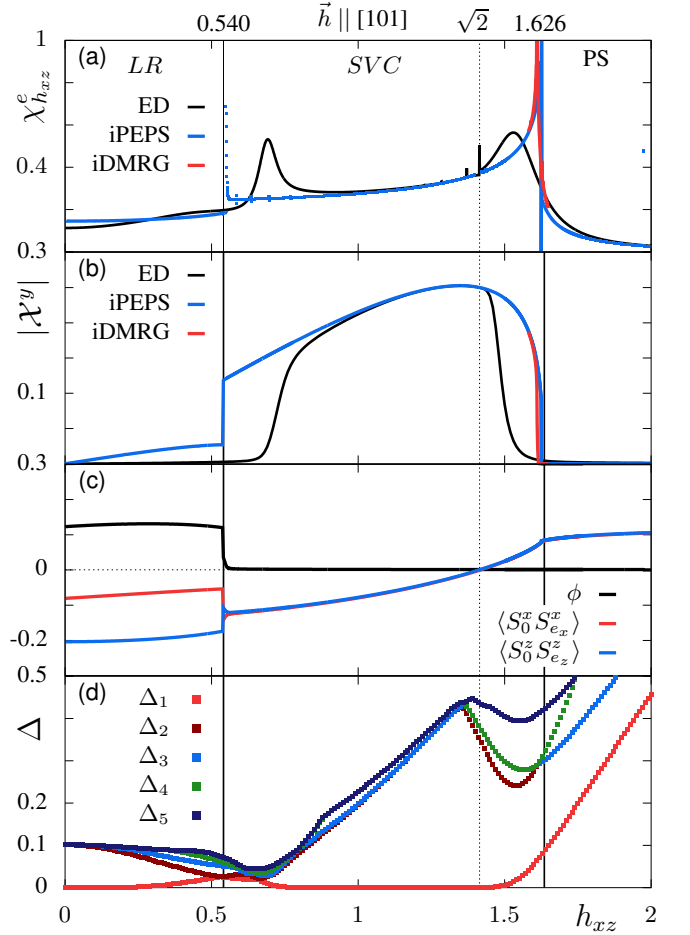

  \centering
  \vspace*{-0.5cm}
  \hspace*{-1.0cm}
  \begin{tikzpicture}
        \node (img1) {\input{full_hxz_egrad2_3.tex}};
        \node [below=of img1, yshift=2.85cm] (img2) {\input{full_hxz_chiral_3.tex}};
        \node [below=of img2, yshift=2.85cm] (img3) {\input{full_hxz_nematic_3.tex}};
        \node [below=of img3, yshift=2.85cm] (img4) {\input{ED_hxz_gaps_3}};
        \node [above=of img1, xshift=-1.3cm, yshift=-1.5cm] (hc11) {\hcone};
        \node [above=of img1, xshift=2.9cm, yshift=-1.5cm] (hc21) {\hctwo};
        \node [above=of img1, xshift=2.0cm, yshift=-1.52cm] (hs1) {$\sqrt{2}$};
        \node [above=of img1, xshift=-3.05cm, yshift=-2.0cm] (lbl1) {\sf (a)};
        \node [above=of img2, xshift=-3.05cm, yshift=-2.0cm] (lbl1) {\sf (b)};
        \node [above=of img1, xshift=-2.35cm, yshift=-2.05cm] (lbl1) {\LF};
        \node [above=of img1, xshift=0.70cm, yshift=-2.05cm] (lbl1) {\SC};
        \node [above=of img1, xshift=3.65cm, yshift=-1.965cm] (lbl1) {PS};
        \node [above=of img1, xshift=0.55cm, yshift=-1.5cm] (lbl1) {$\vec{h}\ ||$ [101]};
        \node [above=of img3, xshift=-3.05cm, yshift=-2.0cm] (lbl1) {\sf (c)};
        \node [above=of img4, xshift=-3.05cm, yshift=-2.0cm] (lbl1) {\sf (d)};
  \end{tikzpicture}
  \vspace*{-1.05cm}
  \caption{Results from ED with PBC on a 4$\times$6 lattice, iDMRG with $L_z$=10, and iPEPS versus field strength, $h_{xz}$, for a field in the [101] direction ($\phi_{xz}=\pi/4$).
  (a) $\chi^e_{h_{xz}}$  ED, iPEPS and iDMRG. (b) $|\mathcal{X}^y|$ from ED
  with a small pinning field 0.005$h_z$ on a single site, iPEPS, and iDMRG. (c) $\phi$ and bond correlations from iPEPS. (d) energy gaps as obtained from ED. Solid vertical lines indicate \hclow=$\hcone$ and \hcup=$\hctwo$ separating the low field $LR$, \SC, and polarized (PS) states. The dotted vertical line indicates the exactly-solvable point, $h^{\star}_{xz}=2SJ\sqrt{2}$.
  } 
  \label{fig:h101}
\end{figure}

{\it Methods:} 
For an in-plane field there is no sign problem and Monte Carlo methods are applicable, but we have found it advantageous to use iPEPS~\cite{Jordan_Cirac_2008, Orus_Vidal_2009, tenes} directly in the thermodynamic limit for the two-dimensional lattice, to obtain high precision results for the field dependent phasediagram of the QCM at zero temperature. For details, see Ref.~\cite{SM}.
In addition, we use exact diagonalization of small clusters, and iDMRG~\cite{White1992a,White1992b,White1993,Schollwock2005,Hallberg2006,Schollwock2011,McCulloch2008,itensor} on infinitely long cylinders in the
$x$-direction, of circumference up to $L_z$=10. 
Typically, we use iDMRG with a bond dimension up to $D\mathord{=} 1000$ and $\epsilon \mathord{=} 10^{-11}$. 
The location of quantum critical points (QCP) are first determined 
from the susceptibility of the ground state energy per spin $e_0$ with respect to a parameter $p$, defined as
    $\chi^e_p=-\frac{\partial^2 e_0 }{\partial p^2}.$
In finite systems, at a quantum critical point,  $\chi^e$ is known to scale as~\cite{venuti2007quantum,Schwandt2009,Albuquerque2010}
  $\chi^e \sim N^{2/\nu-d-z}$, and is therefore likely to diverge at a QCP,
with $\nu$ and $z$ the correlation and dynamical critical exponents and $d$ the spatial dimension. 
Our numerical results are for the $S$=1/2 QCM, and we consider $L_x\mathord{\times} L_z$=$N$ lattices.
In light of our exact solution mentioned previously, and the natural competition between bond-directional ordering of the QCM in zero field, we define the vector bond-chirality 
\begin{align}
\mathcal{X}^y_{\alpha}= \langle\vec{S}_{\bf r} \times \vec{S}_{\bf r+e_{\alpha}}\rangle^y\ \ \ \ \alpha=x,z
\end{align}
along with a nematic order parameter
\begin{align}
    \phi = \langle S^x_{\mathbf{r}} S^x_{\mathbf{r+e_x}}  - S^z_{\mathbf{r}} S^z_{\mathbf{r+e_z}} \rangle .
\end{align}
quantifying the degree of orthogonality and bond-directional alignment of neighboring spins respectively. We have also found it useful to denote the vector chirality averaged over bond directions as $\mathcal{X}^y = \frac{1}{2}\left(\mathcal{X}^y_x +\mathcal{X}^y_z\right)$.

\begin{figure*}[t!]
  \centering
  \hspace*{-0.75cm}
  \begin{tikzpicture}
        \node (img1) {\input{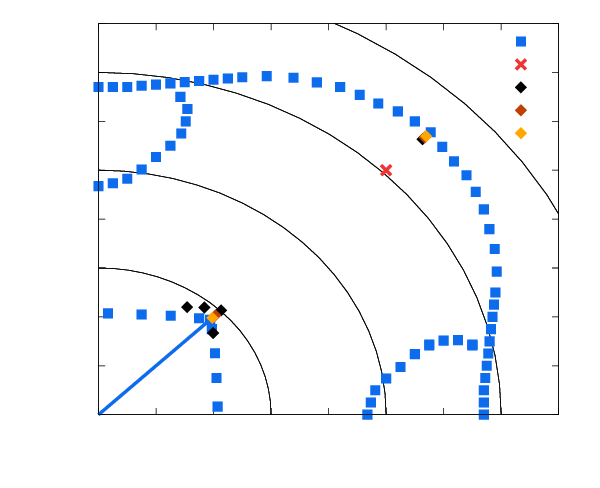}};
        \node [below=of img1, yshift=3.75cm, xshift=-2.7cm] {$L$};
        \node [below=of img1, yshift=3.15cm, xshift=-2.05cm] {$R$};
        \node [below=of img1, yshift=3.10cm, xshift=2.35cm] {\SCX};
        \node [below=of img1, yshift=7.45cm, xshift=-2.80cm] {\SCZ};
        \node [below=of img1, yshift=6.65cm, xshift=1.85cm] (state1) {\color{red} $|P\rangle$};
        \node [below=of img1, yshift=6.0cm, xshift=0.55cm] {\SC};
        \node [right=of img1, xshift=-2cm, yshift=2.15cm] (img2) {\input{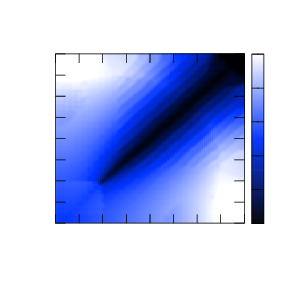}};
        \node [above=of img2, yshift=-2.75cm, xshift=1.0cm] {\color{white} $|\phi|$};
        \node [below=of img2, yshift=2.5cm] (img3) 
        {\input{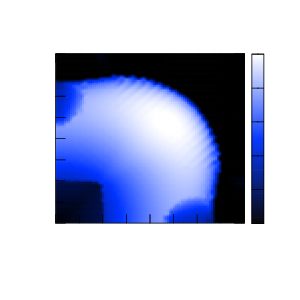}};
        \node [above=of img3, yshift=-2.75cm, xshift=1.0cm] {\color{white} $|\mathcal{X}^y|$};
        \node[right=of img1, yshift=-1.5cm, xshift=4.0cm] {$R$};
        \draw[stealth-, citeblue!90, very thick] (10.0,-2) --   (10.0,-1.7);
        \draw[stealth-, citeblue!90, very thick] (10.3,-2) --   (10.3,-1.7);
        \draw[stealth-, citeblue!90, very thick] (10.6,-2) --   (10.6,-1.7);
        \draw[stealth-, citeblue!90, very thick] (10.9,-2) --   (10.9,-1.7);
        \draw[stealth-, citeblue!90, very thick] (10.0,-2.1) -- (10.0,-2.4);
        \draw[stealth-, citeblue!90, very thick] (10.3,-2.1) -- (10.3,-2.4);
        \draw[stealth-, citeblue!90, very thick] (10.6,-2.1) -- (10.6,-2.4);
        \draw[stealth-, citeblue!90, very thick] (10.9,-2.1) -- (10.9,-2.4);
        \draw[stealth-, citeblue!90, very thick] (10.0,-2.8) -- (10.0,-2.5);
        \draw[stealth-, citeblue!90, very thick] (10.3,-2.8) -- (10.3,-2.5);
        \draw[stealth-, citeblue!90, very thick] (10.6,-2.8) -- (10.6,-2.5);
        \draw[stealth-, citeblue!90, very thick] (10.9,-2.8) -- (10.9,-2.5);
        %
        \node[right=of img1, yshift=0.5cm, xshift=4.0cm] {\SCX};
        \draw[stealth-, citeblue!90, very thick] (10.15,0.15) -- (9.85,0.15);
        \draw[stealth-, citeblue!90, very thick] (10.3,0    ) -- (10.3,0.3);
        \draw[stealth-, citeblue!90, very thick] (10.75,0.15) -- (10.45,0.15);
        \draw[stealth-, citeblue!90, very thick] (10.9,0    ) -- (10.9,0.3);
        \draw[stealth-, citeblue!90, very thick] (10.15,-0.25) -- (9.85,-0.25);
        \draw[stealth-, citeblue!90, very thick] (10.3,-0.1  ) -- (10.3,-0.4);
        \draw[stealth-, citeblue!90, very thick] (10.75,-0.25) -- (10.45,-0.25);
        \draw[stealth-, citeblue!90, very thick] (10.9,-0.1  ) -- (10.9,-0.4);
        \draw[stealth-, citeblue!90, very thick] (10.15,-0.65) -- (9.85,-0.65);
        \draw[stealth-, citeblue!90, very thick] (10.3,-0.8  ) -- (10.3,-0.5);
        \draw[stealth-, citeblue!90, very thick] (10.75,-0.65) -- (10.45,-0.65);
        \draw[stealth-, citeblue!90, very thick] (10.9,-0.8  ) -- (10.9,-0.5);
        \node[right=of img1, yshift=2.5cm, xshift=3.8cm] {\SC};
        \draw[stealth-, citeblue!90, very thick] (10.15,2.15) -- (9.85, 2.15);
        \draw[stealth-, citeblue!90, very thick] (10.3,2.3)   -- (10.3,2.0);
        \draw[stealth-, citeblue!90, very thick] (10.75,2.15) -- (10.45,2.15);
        \draw[stealth-, citeblue!90, very thick] (10.9,2.3)   -- (10.9,2.0);
        \draw[stealth-, citeblue!90, very thick] (10.0,1.9)   -- (10.0,1.6);
        \draw[stealth-, citeblue!90, very thick] (10.45,1.75) -- (10.15,1.75);
        \draw[stealth-, citeblue!90, very thick] (10.6,1.9)   -- (10.6,1.6);
        \draw[stealth-, citeblue!90, very thick] (11.05,1.75) -- (10.75,1.75);
        \draw[stealth-, citeblue!90, very thick] (10.15,1.35) -- (9.85, 1.35);
        \draw[stealth-, citeblue!90, very thick] (10.3,1.5)   -- (10.3,1.2);
        \draw[stealth-, citeblue!90, very thick] (10.75,1.35) -- (10.45,1.35);
        \draw[stealth-, citeblue!90, very thick] (10.9,1.5)   -- (10.9,1.2);
        \node[right=of img1, yshift=-1.5cm, xshift=6.00cm] {$L$};
        \draw[stealth-, citeblue!90, very thick] (12.15,-1.85) --   (11.85,-1.85);
        \draw[stealth-, citeblue!90, very thick] (12.15,-1.85) --   (12.44,-1.85);
        \draw[stealth-, citeblue!90, very thick] (12.75,-1.85) --   (12.46,-1.85);
        \draw[stealth-, citeblue!90, very thick] (12.75,-1.85) --   (13.05,-1.85);
        \draw[stealth-, citeblue!90, very thick] (12.15,-2.25) -- (11.85,-2.25);
        \draw[stealth-, citeblue!90, very thick] (12.15,-2.25) -- (12.44,-2.25);
        \draw[stealth-, citeblue!90, very thick] (12.75,-2.25) -- (12.46,-2.25);
        \draw[stealth-, citeblue!90, very thick] (12.75,-2.25) -- (13.05,-2.25);
        \draw[stealth-, citeblue!90, very thick] (12.15,-2.65) -- (11.85,-2.65);
        \draw[stealth-, citeblue!90, very thick] (12.15,-2.65) -- (12.44,-2.65);
        \draw[stealth-, citeblue!90, very thick] (12.75,-2.65) -- (12.46,-2.65);
        \draw[stealth-, citeblue!90, very thick] (12.75,-2.65) -- (13.05,-2.65);
        %
        \node[right=of img1, yshift=0.5cm, xshift=6.0cm] {\SCZ};
        \draw[stealth-, citeblue!90, very thick] (12.15,0.15) -- (11.85,0.15);
        \draw[stealth-, citeblue!90, very thick] (12.15,0.15) -- (12.44,0.15);
        \draw[stealth-, citeblue!90, very thick] (12.75,0.15) -- (12.46,0.15);
        \draw[stealth-, citeblue!90, very thick] (12.75,0.15) -- (13.05,0.15);
        \draw[stealth-, citeblue!90, very thick] (12.0,-0.1)   -- (12.0,-0.4);
        \draw[stealth-, citeblue!90, very thick] (12.3,-0.1)   -- (12.3,-0.4);
        \draw[stealth-, citeblue!90, very thick] (12.6,-0.1)   -- (12.6,-0.4);
        \draw[stealth-, citeblue!90, very thick] (12.9,-0.1)   -- (12.9,-0.4);
        \draw[stealth-, citeblue!90, very thick] (12.15,-0.65) -- (11.85,-0.65);
        \draw[stealth-, citeblue!90, very thick] (12.15,-0.65) -- (12.44,-0.65);
        \draw[stealth-, citeblue!90, very thick] (12.75,-0.65) -- (12.46,-0.65);
        \draw[stealth-, citeblue!90, very thick] (12.75,-0.65) -- (13.05,-0.65);
        \node[right=of img1, yshift=2.5cm, xshift=5.95cm] {PS};
        \draw[stealth-, citeblue!90, very thick] (12.0,1.5) -- (12.0,1.2);
        \draw[stealth-, citeblue!90, very thick] (12.3,1.5) -- (12.3,1.2);
        \draw[stealth-, citeblue!90, very thick] (12.6,1.5) -- (12.6,1.2);
        \draw[stealth-, citeblue!90, very thick] (12.9,1.5) -- (12.9,1.2);
        \draw[stealth-, citeblue!90, very thick] (12.0,1.9) -- (12.0,1.6);
        \draw[stealth-, citeblue!90, very thick] (12.3,1.9) -- (12.3,1.6);
        \draw[stealth-, citeblue!90, very thick] (12.6,1.9) -- (12.6,1.6);
        \draw[stealth-, citeblue!90, very thick] (12.9,1.9) -- (12.9,1.6);
        \draw[stealth-, citeblue!90, very thick] (12.0,2.3) -- (12.0,2.0);
        \draw[stealth-, citeblue!90, very thick] (12.3,2.3) -- (12.3,2.0);
        \draw[stealth-, citeblue!90, very thick] (12.6,2.3) -- (12.6,2.0);
        \draw[stealth-, citeblue!90, very thick] (12.9,2.3) -- (12.9,2.0);
        \node [above=of img1, xshift=-3.05cm, yshift=-2.1cm] (lbl1) {\sf (a)};
        \node [above=of img2, xshift=-1.25cm, yshift=-2.6cm] (lbl1) {\sf (b)};
        \node [above=of img3, xshift=-1.25cm, yshift=-2.55cm] (lbl1) {\color{white} \sf (c)};
        \node [above=of img2, xshift=3.25cm, yshift=-3.1cm] (lbl1) {\sf (d)};
  \end{tikzpicture}
  \vspace*{-0.35cm}
  \caption{(a) Phase diagram for the quantum compass model under an in-plane field. The phases are labelled as: $z$-oriented stripe (R), $x$-oriented stripe (L), staggered vector chiral (\SC), $z$-chain (\SCX), and $x$-chain (\SCZ). We show iPEPS results (blue squares) and iDMRG (colored diamonds) for $L_z$=6,8,10. Solid black lines are contours of constant field strength. (b) $|\phi|$ and (c) $|\mathcal{X}^y|$ as obtained from iPEPS for an in-plane field. (d) Dominant ordering of states in the labelled phases, the PS state is meant to show alignment in the field direction.
  }
  \label{fig:phase_diagram}
\end{figure*}
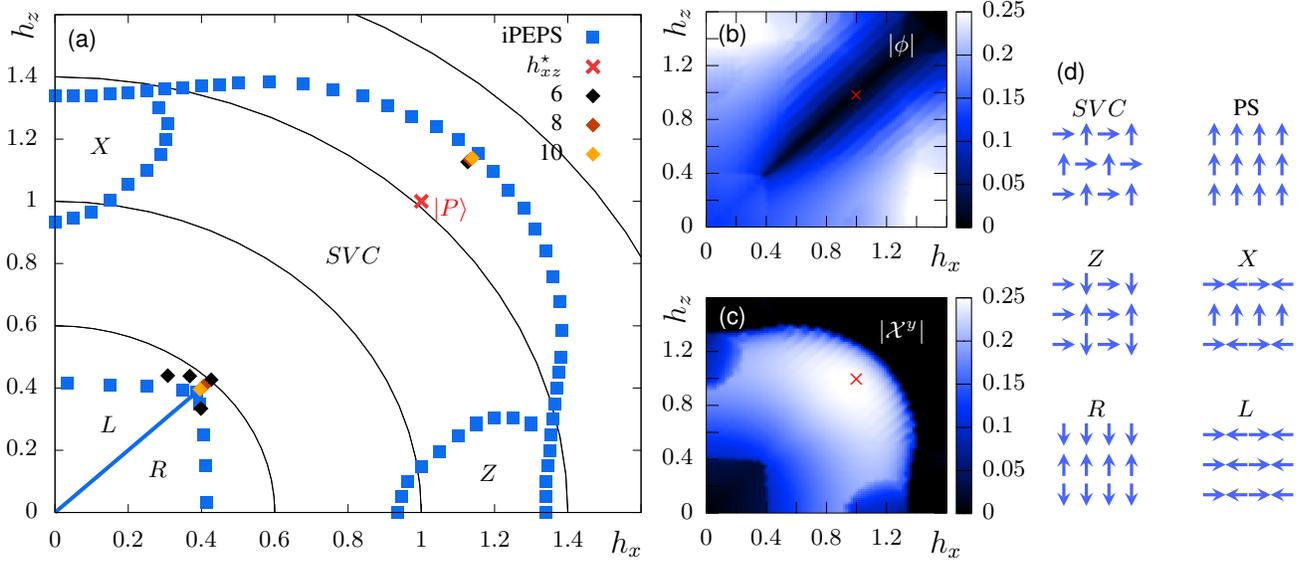

{\it Phases Under [101] Field:} Our iPEPS, iDMRG and ED calculations can clearly distinguish two phase transitions when varying the strength of the in-plane field along the constant angle $\phi_{xy}\mathord{=}\frac{\pi}{4}$ as shown in Fig.\ref{fig:h101}. The high-field phase is a trivial polarized state (PS). Upon lowering the field, at the upper critical field \hcup=\hctwo, the PS transitions into a phase with substantial vector chirality (\SC). This can be seen in 
Fig.\ref{fig:h101}(b), where, at \hcup, $|\mathcal{X}^y|$ increases, seemingly continuously, from zero in the PS, while a divergence in $|\chi^e_{h_{xz}}|$ is observed. Within the \SC\ phase, bond-correlations of the form $\langle S^\alpha S^\alpha\rangle$, with $\alpha\mathord{=}(x,y,z)$ tend to zero as the state approaches the exactly solved states (shown in Fig.~\ref{fig:exact}), at $h^{*}_{xz}\mathord{=}2JS\sqrt{2}$. 
ED results for the gaps, in Fig.\ref{fig:h101}(c), show that the \SC\ phase is gapped with a twofold degenerate ground state.
A second transition into a low-field region with stripe ordering, occurs as the field is lowered below \hclow=\hcone. Within the low field region, the line $h_x\mathord{=}h_z$ for $h_{xz}\mathord{<}h_{xz}^{c1}$ is a first-order critical line, terminating at $h_{xz}^{c1}$, separating phases of $x$-aligned and $z$-aligned stripe states~\cite{SM} that we denote by \LFX\ and \LFZ\ (See Fig.~\ref{fig:phase_diagram}). 
As the field is lowered further to $h_{xz}\mathord{=}0$, we find that the nematic order parameter, shown in Fig.~\ref{fig:h101}(c), saturates to $\phi=0.126$, in agreement with previous quantum Monte Carlo calculations~\cite{Wenzel_Janke_2008}.

{\it Phases Under [100] Field:} Notably, the zero-field QCM has the 1D gauge-like symmetries,
\begin{align}
    P_i = \prod_j S^x_{ie_x + je_z} \ \ \text{and} \ \ Q_i = \prod_j S^z_{je_x + ie_z},
    \label{eq:symmetries}
\end{align}
where the $P_i$ and $Q_i$ are incompatible. Arguments based on symmetry analysis imply that the $S$=$1/2$ QCM ground state is at least 2-fold degenerate~\cite{Doucot_Ioselvich_2005}. However, exact diagonalization calculations indicate that, when $L_x$=$L_z$=$L$, $2\mathord{\times}2^{L}$-$2$ low-energy states collapse onto the $2$-fold ground states exponentially fast with increasing $L$~\cite{Dorier_Becca_2005}, implying an emergent sub-extensive degeneracy in the thermodynamic limit. Following Ref.~\cite{Orus_Vidal_2009_compass}, we label the eigenstates of the $P_i$ and $Q_i$, as $|R\rangle$ and $|L\rangle$, respectively.

We have found that adiabatically evolving the $|R\rangle$ and $|L\rangle$ states under a small [100] field, $h_x$, produces an energy splitting between the two states, with the $|R\rangle$-evolved state, $|R(\vec{h})\rangle$, having lower energy than the $|L\rangle$-evolved state, $|L(\vec{h})\rangle$, for $h_x\mathord{>}h_z$. On the other hand, for a small [001] field, $h_z$, it is $|L(\vec{h})\rangle$ that has the lowest energy.
Consequently, as outlined above, there is a first-order transition between the $|L(\vec{h})\rangle$ and $|R(\vec{h})\rangle$ states
along the line $h_x$=$h_z$~\cite{SM}, reminiscent of the first-order transition studied in Ref.~\cite{Orus_Vidal_2009_compass}.
We find that this line of first-order transitions terminates at the critical point, \hclow=\hcone, discussed in the previous section.
We may then view \hclow\ as a multicritical point since the \LFZ, \LFX\ and \SC\ phase all meet at this point. Furthermore, our calculations indicate that, in zero field, the $|R\rangle$ and $|L\rangle$ states are sub-extensively degenerate, and that these degeneracies are lifted when small finite fields are applied~\cite{SM}.

\begin{figure}[t!]
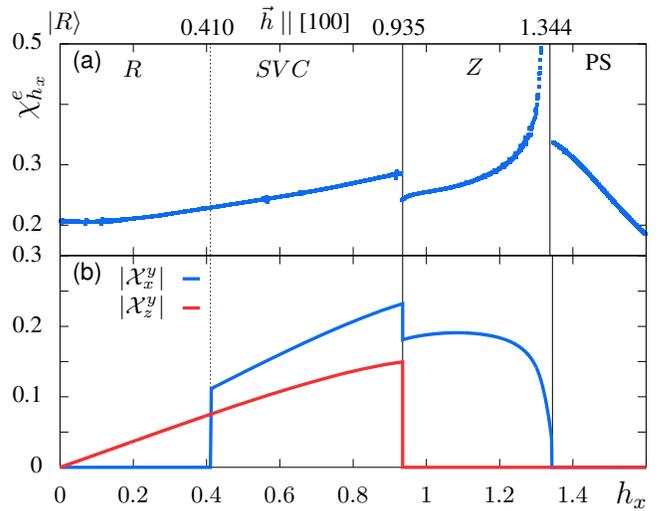

  \centering
  \vspace*{-0.0cm}
  \hspace*{-1.0cm}
  \begin{tikzpicture}
        \node (img1) {\input{full_hx_egrad2_3.tex}};
        \node [below=of img1, yshift=2.85cm] (img2) {\input{full_hx_chiral_3.tex}};
        \node [above=of img1, xshift=1.1cm, yshift=-1.5cm] (hc11) {0.935};
        \node [above=of img1, xshift=3.05cm, yshift=-1.5cm] (hc21) {1.344};
        \node [above=of img1, xshift=-1.45cm, yshift=-1.5cm] (hc21) {0.410};
        \node [above=of img1, xshift=-3.05cm, yshift=-2.0cm] (lbl1) {\sf (a)};
        \node [above=of img2, xshift=-3.05cm, yshift=-2.0cm] (lbl1) {\sf (b)};
        %
        \node [above=of img1, xshift=-2.45cm, yshift=-2.05cm] (lbl1) {$R$};
        \node [above=of img1, xshift=-0.45cm, yshift=-2.05cm] (lbl1) {\SC};
        \node [above=of img1, xshift=2.10cm, yshift=-2.025cm] (lbl1) {\SCX};
        \node [above=of img1, xshift=3.75cm, yshift=-1.965cm] (lbl1) {PS};
        \node [above=of img1, xshift=-0.15cm, yshift=-1.5cm] (lbl1) {$\vec{h}\ ||$ [100]};
        \node [above=of img1, xshift=-3.35cm, yshift=-1.5cm] (lbl1) {$|R\rangle$};
  \end{tikzpicture}
  \vspace*{-1.0cm}
  \caption{(a) $\chi^e_{h_{x}}$ and (b) $|\mathcal{X}^y_x|$ and $|\mathcal{X}^y_z|$ as obtained from iPEPS calculations versus field strength, $h_{x}$, for a field parallel to [100] ($\phi_{xz}=0$). 
  Solid vertical lines indicate $h_{x}^{c1}$=0.935 and $h_{x}^{c2}$=1.344 separating the \SC, \SCX, and polarized (PS) states. A dashed line at $h_x$=0.410 indicates the limiting value of the transition between \LFZ\ and \SC\ phases as $h_z\rightarrow 0$.
  } 
  \label{fig:h100}
\end{figure}

With a field in the [100] direction, the high field PS again undergoes a transition as the field is lowered below a critical field $h_{x}^{c2}$=1.344. However, in this case, the ground state consists approximately of alternating linear domains of field-polarized spins and antiferromagnetically ordered spins perpendicular to the field. The vector chirality is therefore non-zero when evaluated on bonds connected to polarized spins, as shown in Fig.\ref{fig:h100}(b), but only across bonds in the [100] direction. This phase has an interesting interpretation; columns of $x$-polarized spins lowering the energy by aligning with the
field, while columns of $z$-oriented spins form strongly coupled antiferromagnetic Ising chains. Due to the nature of the QCM coupling, the two kinds of columns
are not coupled. This suggests that this state is effectively one-dimensional in nature. For this reason, we refer to this phase as the $z$-chain (\SCX) phase, since the columns of spins polarized along $x$-direction are essentially inert, although their presence effectively eliminates the coupling between the $z$-chains. A sketch of the spin alignments in the \SCX\ phase is shown in Fig.~\ref{fig:phase_diagram}.
As the field is lowered further, a second transition from the {\SCX} phase to the {\SC} phase occurs at $h^{c1}_x$=0.935. 
Finally, as $h_x \rightarrow 0$, the PEPS approaches the $|R\rangle$ state. With $h_z$=0, the transition from the \SC\ phase to the \LFZ\ phase
is not directly visible in $\chi^e_{h_x}$, but the dashed line in Fig.~\ref{fig:h100} indicates the limiting value of the transition between \LFZ\ and \SC\ phases as $h_z\rightarrow 0$, at $h_x=0.410$.
By symmetry of the model, for a field along the $z$-direction an analogous phase, \SCZ, appear along the
$z$ axis. The \SCZ\ phase is dominated by rows of spins coupled by antiferromagnetic $x$-bonds.

{\it Phase Diagram:} We have also analyzed the complete phase diagram for a range of field values, $h_x,h_z\mathord{>}0$. The results of our calculations produce the phase diagram as shown in Fig.~\ref{fig:phase_diagram}. 
The most apparent feature of the phase diagram is the large phase surrounding the point $h^*_{xz}$, where the product states $|P\rangle$, from Fig.~\ref{fig:exact} are exact ground-states. 
As shown in panel (c) of Fig.~\ref{fig:phase_diagram}, the vector chirality, $|\mathcal{X}^y|$, is found to be substantial throughout this phase,
reduced in the \SCX\ and \SCZ\ phases, and approaching zero in the low field \LFX,\LFZ\ regime. The nematic order parameter, $\phi$, is close to zero in the intermediate-field regime for field angles near $\pi/4$ reflecting a lack of spin alignment along bond directions. In the low-field regime, bond-alignment is found to dominate, with $|\mathcal{X}^y|$ taking a value near zero. Remarkably, we find that the transition between the \LFZ\ and
\SC\ phase is almost independent of $h_z$. Likewise, we find the transition between the \LFX\ and \SC\ phase to be independent of $h_x$. The combined
\LFZ\ and \LFX\ phases therefore form a square in the lower left part of the phase diagram, similar to what is seen for the toric code~\cite{Tupitsyn_Stamp_2010}. Even though the field is not applied along an easy axis it is natural to view the \SC\ phase as a spin-flopped phase~\cite{Fisher1974}, and therefore to expect all transitions between the \LFZ, \LFX\, \SCZ\
and \SCX\ to be first order. As it turns out, all our calculations are consistent with this~\cite{SM}. On the other hand, from our calculations, the transition to the PS phase appears to be continuous.

{\it Discussion:}
For the AF QCM we have shown that two exact ground-states exists at the special field value, $h_x^\star$=$h_z^\star$=$2JS$. This special point has a substantial (staggered) vector chirality, $|\mathcal{X}^y|$ and a sizable gap, inducing the \SC\ phase that dominates a large
part of the phase diagram. 
Although our numerical results clearly indicate a sizable gap at $h^\star_{xz}$ within the \SC, establishing a rigorous proof of this gap would be of considerable interest.
Our detailed study of the model under an in-plane magnetic field shows that, aside from the high field PS state, there are five distinct phases in the low to intermediate field regime, the \SC, \SCX, \SCZ, \LFX\  and \LFZ\ phases.  
Excitations in these phases  could be non-trivial. For instance, in zero-field one-dimensional solitonic excitations~\cite{Nussinov_Ortiz_2009} have been noted, and it
is possible that they remain deconfined in the \LFX\  and \LFZ\ phases, as has been observed for the toric code under an in-plane field~\cite{Tupitsyn_Stamp_2010} and the X-cube fracton model~\cite{Devakul_Sondhi_2018}. 
Perhaps the most surprising thing about the phase diagram is the fact that a transition between the \SC\ phase and the PS
phase exists. After all, since the spins are already partly aligned with the field at $h_{XZ}^\star$, one might expect that the PS state could be reached without encountering a
phase transition. The QCM at $h^\star$ would then be at the transition to, or within, the PS phase. But, as we have showed here, $h^\star_{xz}$ is in the distinct \SC\ phase.
In contrast, if we consider 
similar exact product states in Kitaev's honeycomb model (KHCM)~\cite{kitaev2006anyons} with antiferromagnetic couplings, $K_x, K_y, K_z$, then
it is possible to again write the Hamiltonian in the same form as Eq.~(\ref{eq:Hstar}), in terms of a $\H_p$, at the field ($h_x^\star, h_y^\star,h_z^\star)$=$S(K_x,K_y,K_z)$ (note the
factor of 2 difference  with respect to the QCM). However, for the KHCM it is not possible to find an assignment of the $|x\rangle$, $|y\rangle$
and $|z\rangle$ states to the lattice which is an eigenstate of $\H_p$ with eigenvalue 0. However, one might still expect the KHCM at the corresponding
field value of $|h^\star|\mathord{=}KS\sqrt{3}$ for the isotropic model, to be in the same phase as the product state. But, contrary to the QCM, this turns out {\it not} to be the case, since  for that value of $|h^\star|$ the KHCM is known to be in the polarized phase~\cite{sorensen2021prx,hickey2018visons} for a field in the [111] direction. Nevertheless,
it seems possible that similar phases can be found in other materials with predominantly directionally dependent Ising interactions.

\begin{acknowledgments}
We acknowledge the support of
the Natural Sciences and Engineering Research Council of Canada (NSERC) through Discovery
Grant No. RGPIN-2017-05759.
We thank H.-Y. Kee for fruitful discussions. 
This research was enabled in part by support provided by SHARCNET (sharcnet.ca) and the Digital Research Alliance of Canada (alliancecan.ca).
\end{acknowledgments}

\bibliography{references}

\end{document}